\pdfoutput=1
\documentclass[aps, prl, reprint, superscriptaddress, nofootinbib]{revtex4-1}
\usepackage{amsmath, amsfonts, amssymb, hyperref, color, graphicx, xspace, enumitem, slashed}
\usepackage[T1]{fontenc}
\definecolor{nicered}{rgb}{.7,.1,.1}
\definecolor{nicegreen}{rgb}{.1,.5,.1}
\definecolor{darkblue}{rgb}{0,0,.5}
\hypersetup{colorlinks, citecolor=nicegreen, linkcolor=nicered, urlcolor=darkblue}

\begin{document}

\title{Linearly Polarized Gravitational Waves from Bubble Collisions}

\author{Katarina Trailovi\'c}
\email{katarina.trailovic@ijs.si}
\affiliation{Jo\v{z}ef Stefan Institute, Jamova cesta 39, 1000 Ljubljana, Slovenia}
\affiliation{Faculty of Mathematics and Physics, University of Ljubljana, Jadranska ulica 19, 
1000 Ljubljana, Slovenia}

\begin{abstract}
Physics beyond the Standard Model may give rise to first-order phase transitions proceeding via the nucleation of vacuum bubbles, whose subsequent collisions generate gravitational waves (GWs). 
Their detection would open the possibility of investigating the universe in its first instants.
If the transition is slow enough, such that it completes with the nucleation and collision of only two bubbles, the resulting GW signal is linearly polarized. This would give a unique signature for the origin of such a GW signal.
We show that even though such phase transitions would be slow, they still could lie within the detectability range of GW interferometers such as LISA and the Einstein Telescope and the underlying two-bubble origin would be encoded in higher-order polarization statistics. 
\end{abstract}

\maketitle

\section{Introduction}
Future triangular gravitational-wave (GW) detectors such as the Einstein Telescope (ET) and LISA will provide improved sensitivity to GW polarizations, surpassing what is achievable with current networks of ground-based interferometers~\cite{ET:2025xjr,LISA:2024hlh,LIGOScientific:2017ycc,Isi:2017fbj}. Their multiple, non-coaligned arms and long-baseline modulation will enable a more precise reconstruction of the two tensor polarizations predicted in general relativity (GR), as well as stringent tests for additional modes. This capability represents a major advance over existing two- and three-detector networks, for which incomplete baseline coverage and limited signal-to-noise typically restrict the ability to isolate polarization content.

Accurate measurements of GW polarizations open a new observational window to test fundamental physics. On the one hand, they allow powerful constraints on theories that extend GR and predict extra scalar or vector polarizations~\cite{Takeda:2018uai,Hagihara:2019ihn,Yunes:2025xwp,LIGOScientific:2021sio}. On the other hand, even within GR, the polarization state of a GW encodes valuable information about its source. While most astrophysical and cosmological processes generate stochastic backgrounds that are unpolarized on average, notable exceptions exist. For example, axion–gauge-field inflation can produce chiral (circularly polarized) gravitational waves through the amplification of a single helicity mode~\cite{Maleknejad:2016qjz,Caldwell:2017chz}. Such scenarios highlight the possibility that polarization may serve as a diagnostic tool for identifying specific early-Universe mechanisms otherwise inaccessible to direct observation.

In this work, we propose a new mechanism for producing linearly polarized gravitational waves in the early Universe. It has long been understood that collisions of true-vacuum bubbles in first-order phase transitions release substantial energy in the form of gravitational waves, making such transitions among the most promising sources of potentially detectable signals~\cite{Witten:1984rs,Hogan:1986dsh,Kosowsky:1991ua,Kosowsky:1992rz,Kamionkowski:1993fg,Yamada:2025cfr}. 
New physics beyond the Standard Model can give rise to first-order phase transitions in the early Universe. We demonstrate that if such transitions proceed sufficiently slowly to complete through the nucleation and collision of only two vacuum bubbles, the resulting GW signal would exhibit a linear polarization. Using analytical estimates, we further show that slow transitions of this type can occur and successfully complete.
Importantly, this would produce a distinctive and potentially observable signature in future GW detectors, indicative of the unusual dynamics underlying this type of phase transition. Whether this signature can be reconstructed by future detectors depends on the achievable sensitivity to higher-order correlation functions.

\section{Collision of two spherical bubbles}
The collision of two true-vacuum bubbles nucleated during a first-order phase transition generates gravitational radiation \cite{Kosowsky:1991ua}. In the following, we calculate the GW polarization tensor generated by the collision of two spherical bubbles, adapting the analytic formalism of Refs. \cite{Weinberg:1972kfs,Kosowsky:1991ua,Maggiore:2007ulw}.

The linearized Einstein equations in the Lorentz gauge, 
$\partial^\nu \bar{h}_{\mu\nu}=0$, take the form
$\square \bar{h}_{\mu\nu} = -(16\pi G/c^4) T_{\mu\nu}$,
where $\bar{h}_{\mu\nu} = h_{\mu\nu} - \eta_{\mu\nu} h/2$ is the 
trace-reversed metric perturbation. In vacuum, this reduces to 
the wave equation $\square \bar{h}_{\mu\nu} = 0$. Exploiting residual gauge 
freedom, one can impose the transverse-traceless (TT) gauge,
$h^{0\mu}=0, \quad h^i_{\ i}=0, \quad \partial^j h_{ij}=0$,
which isolates the two physical polarization states of the gravitational wave.

For a plane wave $h_{\mu\nu}$ in the Lorentz gauge, the transformation to the TT gauge is obtained 
via the projection
$h_{ij}^\text{TT}=\Lambda_{ij,kl}h_{kl}$,
where $\Lambda_{ij,kl}=P_{ik} P_{jl}-P_{ij}P_{kl}/2$ and $P_{ij}=\delta_{ij}-\hat{k}_i \hat{k}_j$. 
The general plane-wave solution is
$h_{ij}^{\text{TT}} = e_{ij}(\mathbf{k}) e^{ik_\mu x^\mu}$,
where $k^\mu = (\omega, \omega \hat{\mathbf{k}})$ satisfies the null condition 
$k^\mu k_\mu = 0$.

For a propagation direction $\hat{\mathbf{k}}=(\sin\theta,0,\cos\theta)$, 
the transversality condition $\hat{k}^i h_{ij}^{\text{TT}}=0$ implies 
$\sin\theta\, h_{1j} + \cos\theta\, h_{3j}=0$. Combining this with the 
traceless condition and defining $h_{+}\equiv h_{11}$ and $h_{\times}\equiv h_{12}$, 
we obtain the general polarization tensor,
\begin{equation}
e_{ij}(\mathbf{k}) =
\begin{pmatrix}
h_{+} & h_{\times} & -\tan\theta\, h_{+} \\
h_{\times} & -(1+\tan^2\theta)\, h_{+} & -\tan\theta\, h_{\times} \\
-\tan\theta\, h_{+} & -\tan\theta\, h_{\times} & \tan^2\theta\, h_{+}
\end{pmatrix},
\label{eq:ttgeneral}
\end{equation}
representing a GW in TT gauge propagating along 
$\hat{\mathbf{k}}=(\sin\theta,0,\cos\theta)$ with the two independent 
polarization modes $h_{+}$ and $h_{\times}$.

Now, let us calculate $h_{ij}^\text{TT}$ in the case of two spherical bubbles colliding. Far from the source, we have  
\begin{equation}
    h_{ij}^\text{TT}(t,\mathbf{x})=\frac{1}{r} \frac{4 G}{c^5}\Lambda_{ij,kl}(\hat{k})\!\! \int\!\! \frac{d \omega}{2\pi} \tilde{T}_{kl}(\omega, \mathbf{k}) e^{-i\omega (t-r/c)},
    \label{ttfar}
\end{equation}
where $\mathbf{x}=\mathbf{\hat{k}}\cdot r$ and $r\gg d$, with $d$ being the diameter of the source. Also, $\tilde{T}_{kl}$ is the Fourier transform of the stress-energy tensor, whose spatial part takes the form  $T_{i j}=\partial_i \phi \partial_j \phi-\mathcal{L} \delta_{i j}$.
Since $\Lambda_{ij,kl} \delta_{kl}=0$, we will ignore the part proportional to $\delta_{ij}$ and take 
\begin{equation}
    \tilde{T}_{ij}(\omega, \mathbf{k})=\int dt \int d^3 x \ \partial_i \phi \partial_j \phi \ e^{i \omega t-i \mathbf{k}\cdot \mathbf{x}}.
\end{equation}
Considering that our problem is axially symmetric about the $z$ axis connecting the two bubble centers, we can take, without loss of generality, $\mathbf{\hat{k}}=(\sin \theta, 0, \cos \theta)$. Then, axial symmetry and fixing $\hat{k}_y=0$ implies $\tilde{T}_{xy}(\omega, \mathbf{k})=\tilde{T}_{yz}(\omega, \mathbf{k})=0$. Therefore, we obtain
\begin{equation}
\begin{split}
& \Lambda_{ij,kl}(\hat{k}) \tilde{T}_{kl}(\omega, \mathbf{k})\\ &=\begin{pmatrix}
X & 0 & -\tan \theta \ X\\
0& -(1+\tan^2 \theta)\ X & 0\\
-\tan \theta \ X&0& \tan^2 \theta \ X
\end{pmatrix},
\end{split}
\label{eq:poltentwobub}
\end{equation}
with $X=1/2 \cos^2 \theta(\tilde{T}_{xx} \cos^2\theta - \tilde{T}_{yy} + \tilde{T}_{zz} \sin^2 \theta - 
  2 \tilde{T}_{xz} \sin \theta \cos \theta)$. Comparing this matrix with the general parametrization in \eqref{eq:ttgeneral}, we see that $h_{\times}=0$. Hence, in the case of two spherical bubbles colliding, only the $h_{+}$ polarization is generated.

\section{Early Universe Bubbles}
We wish to determine whether a two-bubble completion regime is compatible with successful phase-transition dynamics. To this end we identify the region of parameter space corresponding to an expected bubble multiplicity of order two and verify that transitions in this regime can successfully complete.

 We consider the phase transition to occur after inflation, during the radiation-dominated epoch, where the scale factor evolves as $a \propto t^{1/2}$ and the energy density is given by $\rho = \pi^2 g_\star T^4 / 30$. The Friedmann equation, $H^2 = 8\pi \rho / 3M_p^2$, then implies the scaling relation $t \propto T^{-2}$ between cosmic time and temperature.

In this regime, finite-temperature (thermal) tunneling dominates over quantum tunneling, as the presence of a thermal bath enhances the decay probability of the false vacuum. The decay rate can thus be written as~\cite{Coleman:1977py,Linde:1981zj}
\begin{equation}
\Gamma(T) = T^4 \left( \frac{S_3}{2\pi T} \right)^{3/2} e^{-S_3/T},
\label{eq:thermalgamma}
\end{equation}
where $S_3$ is the three-dimensional Euclidean action of the $O(3)$-symmetric bounce configuration. 

To study the decay rate in a model-independent manner, we parameterize it as $\Gamma(t) = C(t)e^{-A(t)}$. Expanding around the completion time $t_\star$, we obtain $A(t) \simeq A_\star - \beta (t - t_\star)$, where $\beta \simeq \dot{\Gamma}/\Gamma$ characterizes the inverse duration of the phase transition. It is convenient to introduce the dimensionless parameter $\beta_H = \beta / H(t_\star)$, which compares the transition timescale with the Hubble rate.

In contrast to the conventional approach, we do not characterize the phase transition timescale using the standard percolation time. The percolation criterion is motivated by the emergence of a connected, horizon-spanning network of overlapping true-vacuum bubbles and is therefore most appropriate in the many-bubble regime typically considered in studies of fast cosmological phase transitions. In such scenarios, a large number of bubbles nucleate within a Hubble volume, and the formation of a percolating cluster provides a useful proxy for the completion of the transition.

The situation considered here is qualitatively different. Our framework requires an ultra-slow transition in which, on average, only two bubbles nucleate within a Hubble volume before the transition terminates. In this sparse-nucleation regime, the assumptions underlying the percolation criterion are no longer satisfied: there is no large population of bubbles whose overlap can generate a connected percolating network. Instead, the dynamics are governed by the nucleation, expansion, and eventual collision of on average only two bubbles. Consequently, the notion of percolation ceases to be a meaningful indicator of the end of the transition. We therefore define the effective completion time $t_\star$ as the moment at which the false-vacuum survival probability has dropped to the percent level, imposing the stricter condition
$\mathcal{P}_\mathrm{FV}(t_\star) = e^{-I(t_\star)} \simeq 0.01$, where ~\cite{Guth:1979bh,Guth:1981uk,Turner:1992tz} 
\begin{equation}
\begin{split}
    I(t)&=\frac{4\pi}{3}\int_{t_c}^t dt' \Gamma(t') a(t')^3 r(t,t')^3
\end{split}
\end{equation}
is the expected volume of true-vacuum bubbles per unit volume of space at time $t$, $r(t,t')=\int_{t'}^t dt'' v_w/a(t'')$ is the comoving radius at time $t$ of a bubble nucleated at $t'$ propagating with wall velocity $v_w$ and $t_c$ is the critical time at which the true and false vacuum are degenerate.

As we are interested in slow phase transitions that on average nucleate only two bubbles before completion, we must ensure that the total false-vacuum volume, $\mathcal{V}_\text{FV}(t)\propto a(t)^3 \mathcal{P}_\text{FV}(t)$, decreases with time as the transition completes. This requires that the rate of true-vacuum conversion exceeds the dilution from cosmic expansion, implying~\cite{Ellis:2018mja}
\begin{equation}
   \frac{1}{\mathcal{V}_\text{FV}}\frac{d\mathcal{V}_\text{FV}}{dt}=3H-\frac{dI}{dt}<0,
   \label{eq:Vfv}
\end{equation}
at the time of completion.

The expected number of bubbles nucleated up to time $t$ within one Hubble volume is given by~\cite{Athron:2022mmm}
\begin{equation}
N(t) = \frac{4\pi}{3}\int_{t_c}^{t} dt'\, \frac{\Gamma(t') \mathcal{P}_\text{FV}(t')}{H(t')^3}
\end{equation}
where $\mathcal{P}_\text{FV}(t)$ accounts for the false-vacuum fraction to avoid counting bubbles nucleated within already converted regions.

\begin{figure}[htbp]
    \centering
    \includegraphics[width=0.9\linewidth]{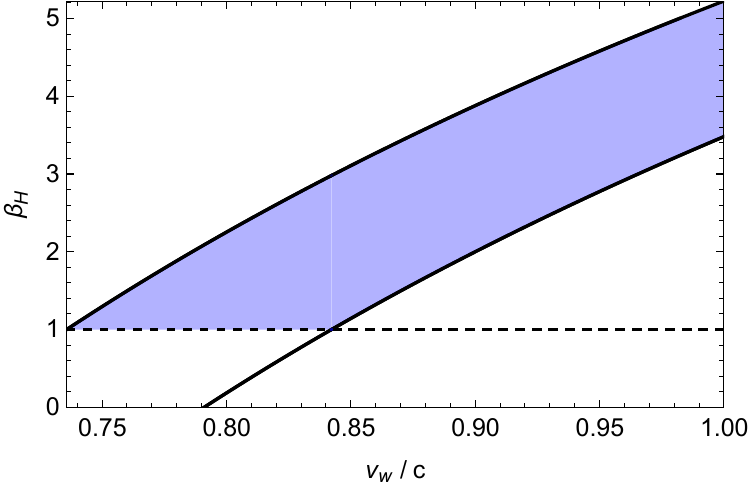}
    \caption{Range of the inverse phase-transition duration, $\beta_H$, for which the expected number of bubbles at completion is two, shown as a function of the bubble wall velocity $v_w$ (purple region). The dashed line indicates $\beta_H = 1$; below this value, nucleation proceeds more slowly than the cosmic expansion, implying that the bubble wall velocity must satisfy $v_w/c > 0.74$.}
    \label{fig:betav}
\end{figure}

In the radiation-dominated era, the completion time can be written as $t_\star = \beta_H / (2\beta)$. Since thermal tunneling dominates in this regime, the prefactor scales as $C(t) \propto T^4 \propto t^{-2}$, and we therefore take $C(t) = \tilde{C}\cdot t^{-2}$ in the following. Furthermore, taking $t_c \simeq 0$\footnote{Setting $t_c = x\, t_\star$ with $x\in[0,1)$, we find that the maximum value occurs at $x=0.026$, corresponding to the mean bubble number reaching $N=3$ already for $\beta_H=1$. Thus, only for $x\in[0,0.026)$ a range of $\beta_H>1$ exists fulfilling $2\le N<3$. This bound is tied to the slow-transition regime: the transition duration admits two consistent estimates, $\tau\sim 1/\beta = 2t_\star/\beta_H$ and $\tau\sim t_\star -t_\text{nuc}\lesssim  t_\star - t_c = t_\star(1-x)$, where $t_\text{nuc}>t_c$ is the time of nucleation, leading to the parametric relation $2/\beta_H \lesssim (1-x)$ and for slow transitions $2/\beta_H\sim\mathcal{O}(1)$, implying $(1-x)\sim \mathcal{O}(1)$ for self-consistency.}
 for analytic simplicity, we obtain
\begin{equation}
\begin{split}
I(t_\star)&=\frac{
32\pi\, B_\star \, e^{-\beta_H/2}
}{
3\, \beta_H^2
} \left(
-4 - 2 e^{\beta_H/2}(-2 + \beta_H)
\right. \\ &\left. + 6\beta_H
+ \sqrt{2\pi}\, (-3 + \beta_H)\, \sqrt{\beta_H}\,
\mathrm{erfi}\!\left( \sqrt{\beta_H/2} \right)
\right)
\label{eq:Itstar}
\end{split}
\end{equation}
and
\begin{equation}
\begin{split}
&N(t_\star)=\frac{32\pi\, B_\star}{3 v_w^3 \beta_H^2} 
\int_{0}^{\beta_H} dx\, x
\exp\Big[\tfrac{x-\beta_H}{2}+\frac{ 32\pi\, B_\star}{
3\beta_H^2
}\\
&  \cdot \Big(
4 - 6x 
+ 2 e^{\frac{x}{2}}(x-2)
+ \sqrt{2\pi x} (3 - x)\,
\mathrm{erfi}\sqrt{\frac{x}{2}}
\Big)\Big],
\label{eq:Ntstar}
\end{split}
\end{equation}
where we defined $B_\star\equiv\tilde{C} t_\star^2 v_w^3$. Imposing $I(t_\star)=4.6$, which encodes a $1\%$ false-vacuum survival probability, and employing Eq.~\eqref{eq:Itstar}, one directly obtains $B_\star$. 
Consequently, Eq.~\eqref{eq:Vfv} depends only on $\beta_H$ and we verified that for any $\beta_H\geq1$ the inequality holds, meaning that completion at this time $t_\star$ is ensured. 
Also, Eq.~\eqref{eq:Ntstar} is now only a function of $\beta_H$ and $v_w$ and we require that the expected number of bubbles at completion satisfies $2 \leq N(t_\star) < 3$. 
This condition yields, for each $v_w$, an allowed range of $\beta_H$. For the case $v_w/c = 1$, we find $3.48 \leq \beta_H < 5.22$ and for smaller wall velocities the allowed ranges of $\beta_H$ are shown in Fig. \ref{fig:betav}.

The mean bubble size $R_\star$ at the time of collision sets the characteristic length scale of the source and thus determines the peak frequency, amplitude, and overall shape of the resulting GW spectrum. In the scenario where two bubbles fill the entire Hubble volume, the mean bubble radius at collision can be approximated as $R_\star H_\star = 0.5$. This estimate can be verified by explicitly computing the mean bubble radius, which at a given time $t$ is given by~\cite{Megevand:2017vtb}
\begin{equation}
    R_\star(t)=\frac{1}{a^3(t) n_B(t)} \int_{t_c}^t dt' \ \Gamma(t') a^3(t') \mathcal{P}_{\text{FV}}(t')R(t,t') ,
    \label{eq:Rmean}
\end{equation}
where $n_B(t)$ is the bubble density, defined as
\begin{equation}
    n_B(t)=\frac{1}{a^3(t)} \int_{t_c}^t dt' \ \Gamma(t') a^3(t') \mathcal{P}_{\text{FV}}(t')
\end{equation}
and $R(t,t')$ is the physical size of a bubble at time $t$ which was nucleated at time $t'$, i.e.\ $ R(t,t')=a(t) r(t,t')$.

\begin{figure}[htbp]
    \centering
    \includegraphics[width=0.9\linewidth]{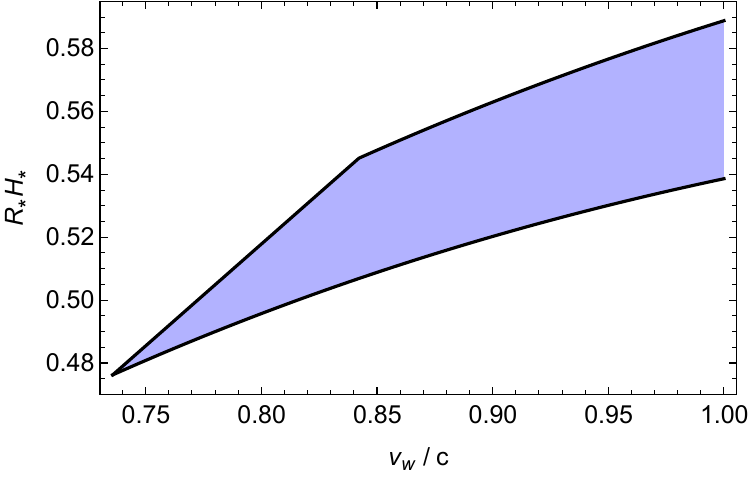}
    \caption{Ratio of the mean bubble size to the Hubble radius at the completion time as a function of the wall velocity $v_w$. The upper (lower) black line corresponds to the lower (upper) bound of $\beta_H$, for which the expected number of bubbles at completion is two. The shaded purple region indicates the allowed values of $R_\star H_\star$, with $\beta_H$ increasing from top to bottom within the permitted range.}
    \label{fig:RH} 
\end{figure}

From the preceding analysis, the ratio of the mean bubble size to the Hubble radius at the completion time $t_\star$ can be expressed as a function of $v_w$ and $\beta_H$, as shown in Fig.~\ref{fig:RH}. The figure indicates that $R_\star H_\star$ at completion is approximately $0.5$, implying that bubble collisions occur at a time close to $t_\star$. This supports the consistency of adopting $R_\star H_\star = 0.5$ as a representative value at the time of collision.

Note that in this regime the nucleated bubbles have radii comparable to the Hubble scale. Consequently, gravitational effects on the bounce action can no longer be neglected. Thus, a first-principles computation of a concrete microscopic model exhibiting such a slow phase transition would require replacing the flat-space Euclidean action $S_3$
  in Eq.~\eqref{eq:thermalgamma} by its gravitationally corrected counterpart, i.e.\ the action obtained from the Coleman–De~Luccia bounce \cite{Coleman:1980aw}. However, our analysis is entirely model-independent, meaning that the parameter $\beta_H$ is treated as an effective quantity, and should therefore be understood as already encoding any gravitational corrections relevant in this slow-transition regime.

\section{GW signal}

Having determined the duration of the phase transition and the mean bubble radius at the time of collision, for a transition that on average completes after the nucleation of two bubbles, our next goal is to estimate the characteristic amplitude and frequency range of the resulting GW signal and assess whether it can plausibly fall within the sensitivity bands of future triangular GW detectors.

There are three processes that contribute to the stochastic GW background: 
\begin{equation}
    \frac{d\Omega_\text{GW} h^2}{d \ln f}\simeq \frac{d \Omega_\phi h^2}{d \ln f} + \frac{d \Omega_\text{sw} h^2}{d \ln f}+ \frac{d \Omega_\text{turb} h^2}{d \ln f},
\end{equation}
the collisions of bubble walls, sound waves and magnetohydrodynamic turbulence in the plasma after collision.

The GW power spectrum today coming from bubble-wall collisions was found by numerical simulations to be the following fitting function  \cite{Cutting:2018tjt}
\begin{equation}
\begin{split}
     \frac{d \Omega_\phi h^2}{d\ln k}\simeq & \ 3.22 \times 10^{-3} (H_\star R_\star)^2 \left(\frac{\kappa_\phi \alpha}{1+\alpha}\right)^2 F_\text{GW}^0\\
     &\times \frac{(a+b)^c \tilde{k}^b k^a}{(b \tilde{k}^{(a+b)/c}+a k^{(a+b)/c})^c},
\end{split}
\end{equation}
where $\alpha=\rho_\text{vac}/\rho_\text{rad}^\star$ is the ratio of the vacuum energy density released during the phase transition to the radiation bath, $\rho_\text{rad}^\star=g_\star \pi^2 T_\star^4/30$ with $g_\star$ the number of relativistic degrees of freedom in the plasma at $T_\star$. $\kappa_\phi=\rho_\phi/\rho_\text{vac}$ is the fraction of vacuum energy density that gets converted into the gradient energy of the scalar, $\tilde{k}=3.2/ R_\star$ is the peak and $a=3$, $b=1.51$, $c=2.18$.
Also, the redshifting factor is given by 
\begin{equation}
\begin{split}
    F_\text{GW}^0&=1.67 \times 10^{-5} \left(\frac{100}{g_\star}\right)^{1/3}.
    \end{split}
\end{equation} 
Thus we obtain
\begin{equation}
\begin{split}
     \frac{d \Omega_\phi h^2}{d\ln f}\simeq & \ 5 \times 10^{-8} (H_\star R_\star)^2 \left(\frac{\kappa_\phi \alpha}{1+\alpha}\right)^2 \left(\frac{100}{g_\star}\right)^{1/3}\\ & \times \frac{(a+b)^c \tilde{f}^b f^a}{(b \tilde{f}^{(a+b)/c}+a f^{(a+b)/c})^c}.
     \end{split}
\end{equation}
The peak frequency generated by the bubble collision today is 
\begin{equation}
    \begin{split}
        & \tilde{f}^0_\phi=\left(\frac{a_\star}{a_0}\right)\frac{3.2}{2\pi R_\star}\\ & \simeq 1.65 \times 10^{-5} \text{Hz} \left(\frac{T_\star}{100 \text{GeV}}\right)\left(\frac{g_\star}{100}\right)^{1/6}\frac{3.2}{2\pi R_\star H_\star}.
    \end{split}
\end{equation}
Although these fitting functions were derived from simulations in the many-bubble regime, Ref. \cite{Cutting:2018tjt} found that the gravitational-wave spectrum depends only weakly on the number of bubbles and that increasing the bubble multiplicity does not significantly alter its shape. We therefore use these fits as indicative estimates for the characteristic frequency and amplitude in the two-bubble regime.

The acoustic sound-wave contribution is given by \cite{Hindmarsh:2017gnf}
\begin{equation}
\begin{split}
    \frac{d \Omega_\text{sw} h^2}{d \ln f}\simeq & 2.061 \ F_\text{GW}^0 \left(\frac{\kappa_{\text{sw}} \alpha}{1+\alpha}\right)^2 (H_\star R_\star) \tilde{\Omega}_\text{GW} \\
    & \times \left(\frac{f}{\tilde{f}_\text{sw}}\right)^3 \left(\frac{7}{4+3(f/\tilde{f}_\text{sw})^2}\right)^{7/2},
    \end{split}
\end{equation}
where $\tilde{\Omega}_\text{GW}=1.2\times 10^{-2}$ from simulations and $\kappa_{\text{sw}}=\rho_{\text{sw}}/\rho_{\text{vac}}$ the fraction of vacuum energy that gets transformed into bulk motion of the fluid. The peak frequency of the sound-wave spectrum today is 
\begin{equation}
    \begin{split}
        \tilde{f}_\text{sw}^0\simeq 4.46 \times 10^{-6} \text{Hz} \left(\frac{T_\star}{100 \text{GeV}}\right)\left(\frac{g_\star}{100}\right)^{1/6} \frac{\beta}{v_w H_\star}\frac{\tilde{k} R_\star}{10},
    \end{split}
\end{equation}
where $\tilde{k}$ is the angular peak frequency of the sound-wave spectrum.

Finally, the contribution from magneto-hydrodynamic turbulence in the plasma can be modeled as \cite{Caprini:2015zlo}
\begin{equation}
    \begin{split}
       \frac{d\Omega_\text{turb} h^2}{d\ln f}
       = &\, 3.35 \times 10^{-4} \left( \frac{H_\star}{\beta} \right)
       \left( \frac{ \kappa_\text{turb} \alpha }{1 + \alpha} \right)^{3/2}
       \left( \frac{100}{g_\star} \right)^{1/3} \\
       & \times v_w
       \frac{ \left( f / \tilde{f}_\text{turb} \right)^3 }
            { \left( 1 + f / \tilde{f}_\text{turb} \right)^{11/3}
              \left( 1 + 8 \pi f / h_\star \right) }, 
    \end{split}
\end{equation}
where $\kappa_\text{turb}=\rho_\text{turb}/\rho_\text{vac}$ and the inverse Hubble time at GW production, redshifted today,
\begin{equation}
    h_\star=1.65\times 10^{-5} \text{Hz}\left(\frac{T_\star}{100\text{GeV}}\right) \left(\frac{g_\star}{100}\right)^{1/6}.
\end{equation}
The turbulence peak frequency today is
\begin{equation}
    \tilde{f}^0_\text{turb}\simeq 2.7\times 10^{-5}\text{Hz}\frac{1}{v_w} \left(\frac{\beta}{H_\star}\right)\left(\frac{T_\star}{100\text{GeV}}\right)\left(\frac{g_\star}{100}\right)^{1/6}.
\end{equation}
\begin{figure}[htbp]
    \centering
    \includegraphics[width=0.99\linewidth]{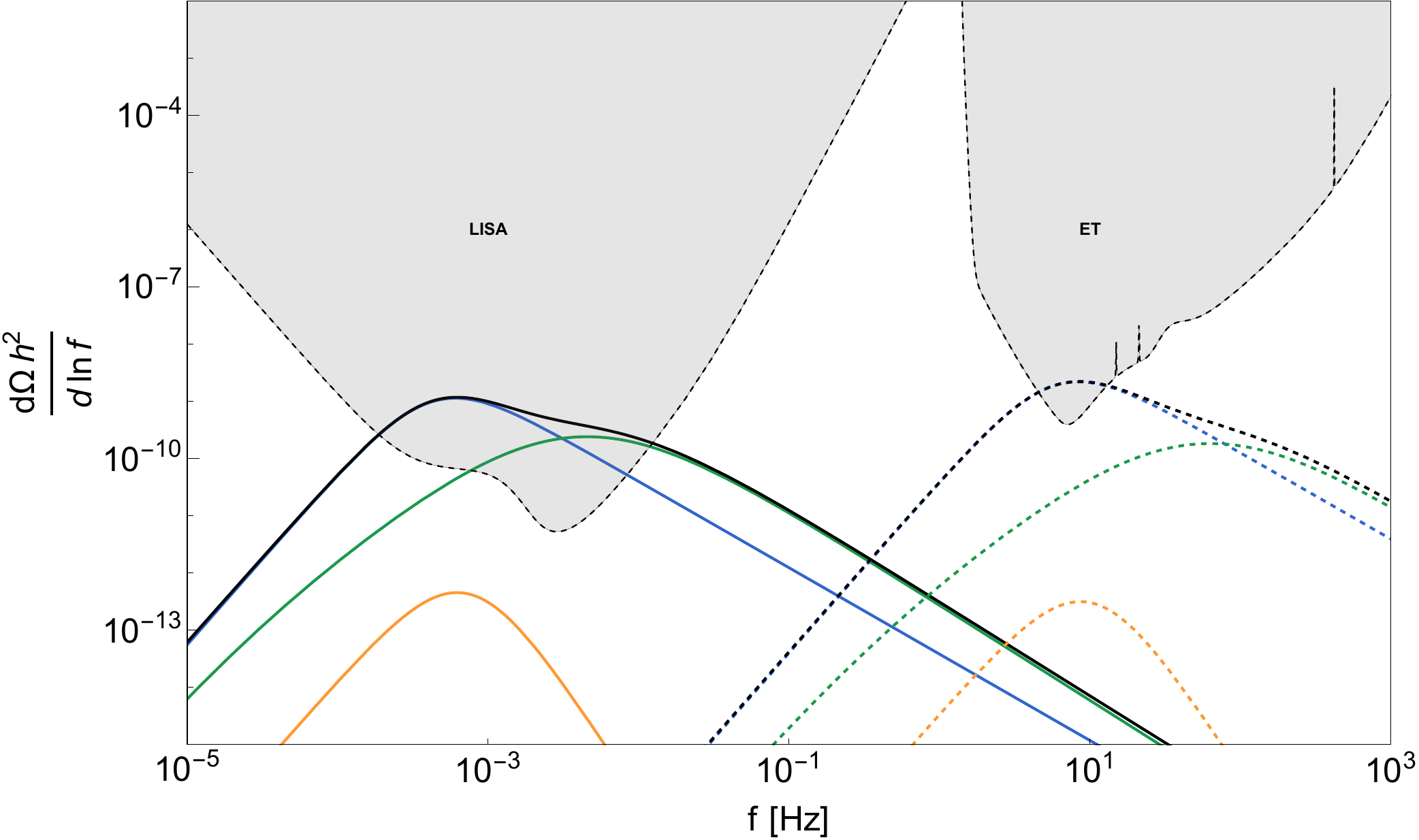}
    \caption{Total stochastic GW background (black) from a radiation-dominated phase transition and its individual contributions from bubble-wall collisions (blue), sound waves (orange), and hydrodynamic turbulence (green), for $\alpha = 0.5$, $T_\star = 3.6 \times 10^3~\text{GeV}$ (solid) and $\alpha = 0.8$, $T_\star = 5.0 \times 10^3~\text{GeV}$ (dashed). The projected sensitivity curves of the future detectors LISA and Einstein Telescope (ET) are shown as grey regions. }
    \label{fig:GWsignal}
\end{figure} 

We assume that the bubbles are in the run-away regime and hence we parametrize \cite{Caprini:2015zlo}
\begin{equation}
    \kappa_\phi=\frac{\alpha-\alpha_\infty}{\alpha}, \ \kappa_\text{sw}=\frac{\alpha_\infty^2}{\alpha (0.73+0.083 \sqrt{\alpha_\infty}+\alpha_\infty)} 
\end{equation}
and $\kappa_\text{turb}=1-\kappa_\phi-\kappa_\text{sw}$, where $\alpha_\infty < \alpha$ corresponds to the onset of the runaway regime, in which the driving pressure from the released vacuum energy surpasses the frictional force from the surrounding plasma, leading to indefinite acceleration of the bubble walls towards the speed of light.
 
In Fig. \ref{fig:GWsignal}, we present the total stochastic GW background coming from a radiation-dominated phase transition along with its individual contributions, computed for the parameters $\alpha = 0.5$ (solid) and $\alpha=0.8$ (dashed), $\alpha_\infty = 0.05$, $T_\star = 3.6 \times 10^3~\text{GeV}$ (solid) and $T_\star = 5.0 \times 10^7~\text{GeV}$ (dashed), $g_\star = 100$, $v_w = 0.95$ and $\tilde{k}R_\star = 10$, with $\beta_H$ chosen at the midpoint of the allowed range.

For this choice of parameters, the estimated GW spectrum overlaps the projected sensitivity bands of LISA and the Einstein Telescope (ET). In particular, the contribution from bubble-wall collisions, corresponding to the linearly polarized component, lies within the projected sensitivity range of LISA and ET. The sound-wave and turbulence contributions, which are expected to be non-polarized, are subdominant and can even fall outside the detectability range. The peak of the bubble-collision component remains within the LISA band for transition temperatures $T_\star$ between $5.5 \times 10^2$ GeV and $1.5 \times 10^5$ GeV and within the ET band for $T_\star$ between $2.5 \times 10^7$ GeV and $1.0 \times 10^8$ GeV .
Varying the bubble-wall velocity $v_w$ does not qualitatively affect this result. Similarly, increasing $\alpha$ within the range $0.5 < \alpha \leq 1$ slightly raises the peak amplitude, but the variation remains within one order of magnitude. In contrast, smaller values of $\alpha$ lead to a faster suppression of the peak amplitude, potentially causing the signal to drop below LISA’s or ET's sensitivity range.
Changing $\alpha_\infty$ modifies the relative contributions from bubble-wall collisions, sound waves, and turbulence.

We also note that the axial symmetry of the two-bubble collision geometry enforces an angular dependence in the GW amplitude: it vanishes along the axis connecting the two bubble centers, peaks in the orthogonal plane, and remains nonzero at intermediate angles~\cite{Kosowsky:1991ua}. Since exact alignment occupies zero solid angle on the sphere, it occurs with probability zero; thus, a null signal is not realized for any generic observation direction. This projection factor can be absorbed into the phase-transition strength parameter, $\alpha$, without loss of generality.

For a supercooled scenario ($\alpha > 1$), the resulting GW signal would be even stronger~\cite{Levi:2022bzt,Ellis:2020nnr,Ellis:2018mja,Coleman:1973jx,Gildener:1976ih,Witten:1980ez,Hambye:2013dgv,Iso:2017uuu,Azatov:2019png,Randall:2006py,Nardini:2007me,Konstandin:2011dr}. However, as discussed in the Appendix, it is unlikely that a supercooled phase transition can complete through the nucleation of only two bubbles; in this regime, successful completion typically requires the nucleation of many bubbles. Nevertheless, in finely tuned scenarios it may still be possible for the transition to complete with only two.

\section{Observable Signatures of Linearized Polarization}
In the previous section, we showed that the gravitational-wave signal produced by bubble collisions in phase transitions with an expected bubble multiplicity of order two at completion can lie within the sensitivity range of future detectors. The distinctive feature of this scenario is that individual two-bubble collisions generate linearly polarized gravitational waves. We therefore investigate in the following what observable signatures of this underlying polarization structure may survive in the stochastic gravitational-wave background observed today.

\subsection{Stokes Parameters}
It is important to emphasize that the two tensor polarizations are frame dependent: a rotation around the propagation axis mixes $h_{+}(t)$ and $h_{\times}(t)$. A frame-independent characterization of the polarization state is therefore obtained in terms of the Stokes parameters~\cite{Seto:2008sr,Gubitosi:2016yoq,Kato:2015bye,Conneely:2018wis}:
\begin{equation}
\begin{split}
    U(\omega, \mathbf{k})&=-2 \langle \Re \ [h_{+}(\omega, \mathbf{k}) h_{\times}^*(\omega, \mathbf{k})]\rangle, \\ V (\omega, \mathbf{k})&=-2 \langle \Im \ [h_{+}(\omega, \mathbf{k}) h_{\times}^*(\omega, \mathbf{k})]\rangle,\\
    I(\omega, \mathbf{k})&=\langle |h_{+}(\omega, \mathbf{k})|^2+|h_{\times}(\omega, \mathbf{k})|^2\rangle, \\ Q(\omega, \mathbf{k})&=\langle |h_{+}(\omega, \mathbf{k})|^2-|h_{\times}(\omega, \mathbf{k})|^2\rangle.
    \end{split}
\end{equation}
where the brackets denote an ensemble average over realizations of the gravitational wave signal.
It is convenient to introduce the helicity basis $h_{R/L} = h_{+} \pm i h_{\times}$, which under a rotation by an angle $\theta$ about the propagation axis transform as $h_{R/L} \rightarrow e^{\mp 2 i \theta} h_{R/L}$, corresponding to helicities $\pm 2$. Under such rotations, $I$ and $V$ are invariant, whereas $Q$ and $U$ mix. The combination $\sqrt{Q^2 + U^2}$, however, is rotationally invariant. A gravitational wave signal is linearly polarized if there exists a polarization basis in which $U = V = 0$ and $I = Q$, which equivalently implies $V=0$ and $\sqrt{Q^2+U^2}=I$ in any frame. The corresponding degree of polarization is
\begin{equation}
    P=\frac{\sqrt{Q^2+U^2+V^2}}{I},
\end{equation}
with $P=1$ for a fully polarized signal.

However, in a realistic cosmological setting our present Hubble volume contains many causally disconnected Hubble patches from the epoch of the phase transition and each patch can contain bubble configurations with different orientations.
For this let us work in the observer's frame, with the line of sight
along $\hat{k} = \hat{z}$.
Because the two-bubble collision is axially symmetric around the collision axis $\hat{n}$,
the stress-energy tensor in the observer's frame takes the form
\begin{equation}
  \tilde{T}_{ij}(\omega, \mathbf{k}) = T_\perp(\omega, \mathbf{k})\,\delta_{ij} + (T_\parallel(\omega, \mathbf{k}) - T_\perp(\omega, \mathbf{k}))\,\hat{n}_i\hat{n}_j, 
\end{equation}
where $T_\parallel$ ($T_\perp$) is the component along (transverse to)
the collision axis.
The collision axis of the two bubbles is a general unit vector
\begin{equation}
  \hat n= (\sin\alpha\cos\psi,\; \sin\alpha\sin\psi,\; \cos\alpha),
\end{equation}
where $\alpha$ is the polar angle of the collision axis from the line of
sight, and $\psi$ is its azimuthal angle in the transverse plane.
Applying the TT projection $\Lambda_{ij,kl}(\hat{z})$ with $\hat k = \hat{z}$, the
result for $i,j \in \{x,y\}$ is
\begin{equation}
  h_{ij}^{\rm TT}(\omega, \mathbf{k}) = \frac{1}{2}(T_\parallel - T_\perp)\sin^2\!\alpha
  \begin{pmatrix} \cos 2\psi & \sin 2\psi \\ \sin 2\psi & -\cos 2\psi \end{pmatrix},
\end{equation}
Therefore, the two polarization amplitudes are
\begin{equation}
\begin{split}
 h_+ (\omega, \mathbf{k})&= A(\omega, \mathbf{k})\sin^2\!\alpha\,\cos(2\psi), \\
    h_\times (\omega, \mathbf{k}) &=  A(\omega, \mathbf{k})\sin^2\!\alpha\,\sin(2\psi),
    \end{split}
  \label{eq:polarizations}
\end{equation}
where $A \equiv \tfrac{1}{2}(T_\parallel - T_\perp)$ is the intrinsic amplitude.

Across the ensemble of Hubble patches, the collision axis $\hat{n}$ takes
a random orientation. Therefore we average isotopically over all possible orientations\footnote{Note that we do not average over $A(\omega, \mathbf{k})$ since it is not a random variable, it depends on the parameters of the phase transition.}, i.e.\ \begin{equation}
    \langle X \rangle\equiv\frac{1}{4\pi}\int_0^{2\pi} d\psi \int_0^\pi d\alpha \, \sin \alpha \, X,
    \label{expval}
\end{equation} which gives $Q=U=V=0$, while $I\neq 0$. Hence $P=0$ and thus, while each realization is fully linearly polarized, the ensemble-averaged Stokes parameters vanish, implying that the mean signal is unpolarized.

Therefore, the linear polarization of individual realizations is not directly observable through the ensemble-averaged Stokes parameters. To determine whether the underlying polarization structure leaves an observable imprint, one must instead consider higher-order correlation functions, which provide a more suitable probe of the underlying source geometry than the ensemble-averaged Stokes parameters.

\subsection{Non-Gaussianity}

The underlying polarization structure gets revealed when considering $2$-point and $4$-point correlation functions. Namely, for a standard Gaussian stochastic GW background, the
amplitudes ${h}_+$ and ${h}_\times$ are independent complex Gaussian random variables, $h_{+/\times}\sim \mathcal{C} \mathcal{N}(0,\sigma^2)$. Thus we have $Q=V=U=0$, $I\neq 0$ and the Wick factorization
\begin{equation}
    \langle h_i h_j h_k^* h_l^*\rangle=\langle h_i h_j \rangle \langle h_k^* h_l^*\rangle+ \langle h_i h_k^* \rangle \langle h_j h_l^*\rangle+ \langle h_i h_l^* \rangle \langle h_j h_k^*\rangle
\end{equation}
for $i,j,k,l \in \{+,\times\}$. Hence we can define
\begin{equation}
  \kappa_a\equiv\frac{ \langle|h_a(\omega,\mathbf{k})|^4\rangle}{\langle h_a(\omega,\mathbf{k})^2\rangle \langle h_a^*(\omega,\mathbf{k})^2\rangle+2\langle|h_a(\omega,\mathbf{k})|^2\rangle^2}
\end{equation}
with $a\in \{+,\times\}$ and
\begin{equation}
\begin{split}
    &\kappa_{+\times}\equiv \langle|h_+(\omega,\mathbf{k})|^2 |h_\times(\omega,\mathbf{k})|^2\rangle \cdot(|\langle h_+(\omega,\mathbf{k})h_\times(\omega,\mathbf{k})\rangle|^2\\ &+|\langle h_+(\omega,\mathbf{k})h_\times^*(\omega,\mathbf{k})\rangle|^2 +\langle|h_+(\omega,\mathbf{k})|^2\rangle \langle|h_\times(\omega,\mathbf{k})|^2\rangle)^{-1},
    \end{split}
\end{equation}
where in the case of complex Gaussian random variables we have $\kappa_+=\kappa_\times=\kappa_{+ \times}=1$.

For the ensemble of Hubble patches with two bubble collisions, performing an average over random collision axes orientation, i.e.\ using \eqref{expval}, however we obtain $\kappa_+=\kappa_\times=\kappa_{+ \times}=5/7$, indicating that the gravitational wave signal is non-Gaussian.

Hence, although the orientation-averaged source population has no preferred polarization direction ($P=0$), the fact that each realization is fully linearly polarized leaves a non-Gaussian imprint in higher-order statistics.
The vanishing of $V$ confirms the absence of circular polarization, while $\kappa_+=\kappa_\times=\kappa_{+\times}= 5/7$ encodes the underlying linear nature of the source.

So far, we only considered the intrinsic source-population statistics of the ensemble of randomly oriented two-bubble collisions, which do not yet include the effect of summing many statistically independent signals coming from different Hubble patches in the observed stochastic background. To see this, let
\begin{equation}
H_a = \sum_{i=1}^{N_{\rm eff}} h_{a,i},
\qquad
a\in\{+,\times\},
\end{equation}
where the $h_{a,i}$ are independent realizations drawn from the same two-bubble orientation ensemble and $N_{\rm eff}$ is the effective number of Hubble patches contributing to the signal. For $h_{a,i}$ identical and independent with vanishing mean, one finds
\begin{equation}
\langle H_a^2\rangle \langle H_a^{* 2}\rangle+\langle |H_a|^2\rangle=N_{\rm eff}^2(\langle h_a^2\rangle \langle h_a^{* 2}\rangle+2\langle |h_a|^2\rangle^2),
\end{equation}
 while
 \begin{equation}
\begin{split}
\frac{\langle |H_a|^4\rangle}{ N_{\rm eff}}
&=
\langle |h_a|^4 \rangle
+
( N_{\rm eff}-1)
\left[
\langle h_a^2 \rangle
\langle h_a^{*2} \rangle
+
2\langle |h_a|^2 \rangle^2
\right]\\
&=(\kappa_a+N_{\rm eff} -1)\left[
\langle h_a^2 \rangle
\langle h_a^{*2} \rangle
+
2\langle |h_a|^2 \rangle^2
\right]
\end{split}
\end{equation}
Therefore, the normalized fourth-order statistic of the summed signal becomes
\begin{equation}
\kappa_a^{\rm obs}
=
1+
\frac{\kappa_a-1}
     {N_{\rm eff}}.
\end{equation}
Using $\kappa_a=5/7$, one obtains
\begin{equation}
\kappa_a^{\rm obs}
=
1-\frac{2}{7N_{\rm eff}}.
\end{equation}
Thus, the value $5/7$ corresponds to the intrinsic two-bubble contribution, or to the limiting case in which a given frequency bin is dominated by a single effective realization. In the opposite limit $N_{\rm eff}\gg1$, the observed statistic approaches the Gaussian value $\kappa_a^{\rm obs}=1$, as expected from the central-limit theorem. For $N_{\rm eff}\sim 10$, we get $\kappa_a^{\rm obs}=0.97$, which is still below the Gaussian value. 

Consequently, the non-Gaussian signature becomes progressively diluted as the number of effective Hubble patches contributing to the signal increases. The observability of this feature therefore depends on the epoch of the phase transition: transitions occurring at very early times are expected to involve a larger number of independent contributing patches, suppressing the effect, whereas transitions occurring at later times may retain a potentially observable non-Gaussian imprint.

The detectability of this signal also depends on the precision with which the corresponding four- and two-point correlation functions can be measured. A quantitative assessment of the prospects for reconstructing such higher-order statistics with future gravitational-wave detectors is beyond the scope of the present work

\subsection{Effect of Bubble-Number Fluctuations}

The analysis above focused on the idealized situation in which each Hubble patch completes the phase transition through the nucleation and collision of exactly two bubbles. In practice, however, the condition $N(t_\star)\simeq 2$ refers to the expected number of bubbles per Hubble volume. Since bubble nucleation is a stochastic process, the actual number of bubbles in a given patch fluctuates around the mean. Consequently, even when the average bubble number is close to two, a fraction of Hubble patches will contain three or more bubbles.

 Assuming that the nucleation of bubbles is Poisson distributed, the probability to nucleate $n$ bubbles is 
\begin{equation}
    p_n=\frac{N^n e^{- N}}{n!},
\end{equation}
where $N$ is the expected number of bubbles. For our scenario, where we consider very slow phase transitions, we want to take $2\leq N <3$. This gives a probability of nucleating two bubbles between $0.27\geq p_2 > 0.22$, which is already a non-negligible amount of Hubble patches. The special polarization properties derived in previous sections apply strictly to the $n=2$ component, while for $n>2$, the collision geometry is generically more complicated and the resulting GW signal is not expected to exhibit the same fully linearly polarized structure.

Nevertheless, the contribution from two-bubble patches need not be completely washed out by the presence of higher-multiplicity regions. The characteristic bubble size increases as the number of bubbles per Hubble volume decreases. In the two-bubble regime considered here, the mean bubble radius at collision satisfies $R_\star H_\star \sim 0.5$, corresponding to bubbles that occupy a substantial fraction of the Hubble volume. Since the characteristic GW frequency scales as
$
f_{\rm peak}\propto R_\star^{-1}$,
while the bubble-collision contribution scales approximately as $
\Omega_\phi \propto (H_\star R_\star)^2$,
the signal from two-bubble patches is expected to peak at lower frequencies and may possess a larger amplitude than the contribution from regions containing many smaller bubbles.

As a result, the stochastic background may contain a spectrally distinct component associated with the largest bubbles exhibiting the non-Gaussianity discussed above. Determining how these statistics are reconstructed from an observed stochastic background requires a dedicated analysis, which lies beyond the scope of the present work.

\section{Conclusions}
We have identified a region of parameter space for first-order phase transitions in a radiation-dominated Universe in which the expected bubble multiplicity at completion is of order two.
By determining the corresponding range of $\beta_H$, we showed that transitions in this regime can satisfy the conditions required for successful completion, indicating that they are dynamically viable. 
In this regime, the gravitational-wave signal is dominated by two-bubble collisions and exhibits a linear polarization, providing a distinctive signature that differentiates it from other known cosmological gravitational-wave sources. The observation of such a signal, or of the associated higher-order statistical signatures, would offer a unique probe of the dynamics of phase transitions in the early Universe.

Although the GWs produced by two-bubble collisions originate from slow first-order phase transitions, we have shown that the resulting signals can lie within the sensitivity range of future gravitational-wave observatories such as LISA and the Einstein Telescope. Establishing the polarization properties of such a signal, however, would require the measurement of its 4-point and 2-point correlation functions in order to identify non-Gaussianties. Whether these higher-order statistics can be reconstructed with sufficient precision by future detectors remains an open question. We leave a detailed assessment of their observability and reconstruction prospects to future work.

Finally, it would be valuable to complement these analytical results with dedicated numerical simulations. Also, it would be interesting to construct explicit model frameworks capable of realizing such slow first-order phase transitions.

\section{Acknowledgment}
I thank Diego Redigolo for pointing me to this question and for valuable discussions. I also thank Lorenzo Ubaldi, Miha Nemev\v sek, Alberto Mariotti and Miguel Vanvlasselaer for useful discussions, and Lorenzo Ubaldi in particular for comments on the draft. I further thank Toby Opferkuch for providing the data of the projected noise curves for the future experiments LISA and ET.
This work was supported by the Slovenian Research Agency (research core funding No. P1-0035).

\bibliographystyle{apsrev4-2}
\bibliography{references}

\section{Appendix}
In this Appendix, we wish to determine whether a two-bubble completion regime is compatible with successful phase-transition dynamics for the case of supercooled phase transitions.

In a first-order phase transition, the system may remain temporarily trapped in a metastable false vacuum even after the temperature drops below the critical temperature $T_c$, where the true and false vacuum are degenerate. When this occurs, the Universe continues to cool while the transition to the true vacuum is delayed, leading to an accumulation of vacuum energy in the false vacuum.
This stage, known as supercooling, is characterized by an energy density dominated by the nearly constant false-vacuum component, causing the cosmic expansion to become approximately exponential~\cite{Levi:2022bzt,Ellis:2020nnr,Ellis:2018mja,Coleman:1973jx,Gildener:1976ih,Witten:1980ez,Hambye:2013dgv,Iso:2017uuu,Azatov:2019png,Randall:2006py,Nardini:2007me,Konstandin:2011dr,Yamada:2025hfs}. The resulting behaviour corresponds to a brief, secondary period of inflation preceding the completion of the phase transition.
Thus, the scale factor in this regime is $a(t)\propto e^{H t}$, where the Hubble rate $H$ is approximately constant.
The decay of the false vacuum during supercooling can proceed either through thermal transitions, described by $\Gamma_3(T)$ in Eq.~\eqref{eq:thermalgamma}, or through quantum tunneling, with decay rate $\Gamma_4=R_0^{-4}(S_4/2\pi)^2 e^{-S_4}$, where $S_4$ is the four-dimensional Euclidean action of the $O(4)$-symmetric bounce solution and $R_0$ denotes the bubble radius at nucleation. As before, we study the decay rate in a model-independent manner, taking $C(t) = \tilde{C}\cdot T^4$ when $\Gamma_3(T)$ dominates and $C(t) = \tilde{C}\equiv \text{constant}$ when $\Gamma_4$ dominates.
We define the total number of e-folds of expansion between the critical time $t_c$ and a later time $t$ as
\begin{equation}
    N_\text{tot}(t) = \int_{t_c}^{t} dt'\, H(t') \simeq H (t - t_c).
\end{equation}
Since the temperature redshifts inversely with the scale factor, $T \propto a(t)^{-1} \propto e^{-H t}$,
the number of e-folds between temperatures $T_c$ and $T$ can be expressed as $N_\text{tot}(T) = \ln(T_c/T)$.
This quantity measures the total expansion of the Universe during the supercooled stage, before the phase transition completes.

We first consider the case in which quantum tunneling dominates the decay rate. In this regime, we obtain
\begin{equation}
\begin{split}
  &I(t)=\frac{
4 \pi  \tilde{C} v_w^3}{
3 H^4 \beta_H 
}
\Big(\frac{
6
+ 2 \beta_H e^{- \beta_H N_\text{tot}^\star-3 N_\text{tot}}}{ (1 + \beta_H)(2 + \beta_H)(3 + \beta_H)}- e^{-\beta_H N_\text{tot}^\star }\\ &
+ 
\frac{
\beta_H^2
- 3 e^{ N_\text{tot}} \beta_H (1 + \beta_H)
+ 3 e^{2  N_\text{tot}} \beta_H (2 + \beta_H)
}{e^{ \beta_H N_\text{tot}^\star +3 N_\text{tot}}(1 + \beta_H)(2 + \beta_H)}
\Big),
\label{eq:Iscquant}
\end{split}
\end{equation}
where we define $N_\text{tot}^\star\equiv N_\text{tot}(t_\star)$ and $N_\text{tot}\equiv N_\text{tot}(t)$.
The corresponding expected number of bubbles at completion is given by
\begin{equation}
    N(t_\star)=\frac{4\pi \tilde{C}}{3 H^4}\int_0^{N_\text{tot}^\star} dN_\text{tot} f(N_\text{tot}) e^{\beta_H(N_\text{tot}-N_\text{tot}^\star)}e^{-I(t)},
    \label{eq:Nsc}
\end{equation}
where $f(N_\text{tot})=1$ for quantum transitions and $f(N_\text{tot})=\exp(-4 N_\text{tot})$ for thermal transitions, coming form the respective prefactor of the decay rate. Solving $I(t_\star) = 4.6$ fixes $\tilde{C}$, so that Eq.~\eqref{eq:Nsc} depends only on $\beta_H$ and $N_\text{tot}^\star$.
We have verified that condition~\eqref{eq:Vfv} is satisfied for any $\beta_H \geq 1$ and $N_\text{tot} > 0$.
Moreover, we find that for all $\beta_H \geq 1$ and $N_\text{tot} > 0$, $N(t_\star) > 3$.
Hence, if quantum tunneling dominates, the phase transition is expected to always completes with the nucleation of at least three bubbles and the resulting GW signal is not linearly polarized.

\begin{figure}[t]
    \centering
    \includegraphics[width=0.9\linewidth]{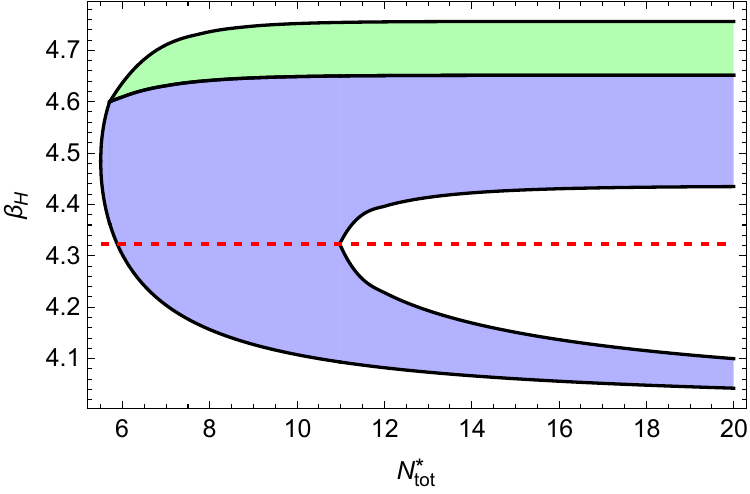}
    \caption{Parameter space of $\beta_H$ and $N_\text{tot}^\star$ for which the expected number of bubbles per Hubble volume at completion satisfies $2 \leq N(t_\star) < 3$ in a supercooled phase transition. The green region corresponds to cases where the completion time is determined by $I(t_\star) = 4.6$, while the purple region denotes the regime where it is set by $(3H)^{-1}dI/dt_{|_{t=t_\star}}=1$. The dashed red line marks $R_\star H_\star=1.8$.}
    \label{fig:betaNtotsc} 
\end{figure}

For the case dominated by thermal tunneling, we find
\begin{equation}
\begin{split}
&I(t)=\frac{4\pi\tilde{C}v_w^3}{3H^4 e^{\beta_H N_\text{tot}^\star}}
\Big(
\frac{6 e^{N_\text{tot} ( \beta_H-4)}
}{
( \beta_H-4 )( \beta_H-3 )( \beta_H-2)(\beta_H-1 )
}\\&
+ \frac{e^{-3 N_\text{tot}}}{( \beta_H-1 )}
- \frac{3 e^{-2 N_\text{tot}}}{(\beta_H-2 )}
+ \frac{3 e^{-N_\text{tot}}}{(\beta_H-3 )}
- \frac{1}{( \beta_H-4 )}
\Big)
\end{split}
\end{equation}
where $\beta_H\in\{1,2,3,4\}$ are removable singularities. In the following analysis, we take $v_w / c = 1$, as bubble walls in the supercooled regime are expected to approach relativistic velocities, being only weakly affected by plasma friction. We find that there exists a function $\bar{\beta}_H(N_\text{tot}^\star)$ such that, for $\beta_H \geq \bar{\beta}_H(N_\text{tot}^\star)$, condition~\eqref{eq:Vfv} is satisfied when $I(t_\star) = 4.6$. For any $N_\text{tot}^\star > 5.7$, there exists a range of $\beta_H$ with $\bar{\beta}_H(N_\text{tot}^\star)$ being the lower bound in which the expected number of bubbles per Hubble volume at completion satisfies $2 \leq N(t_\star) < 3$ (green region in Fig.~\ref{fig:betaNtotsc}).
For $1 \leq \beta_H < \bar{\beta}_H(N_\text{tot}^\star)$, the false-vacuum volume is still non-decreasing when $\mathcal{P}_\text{FV}\simeq 0.01$, therefore we define the completion time $t_\star$ in this regime to correspond to the point at which condition~\eqref{eq:Vfv} is first satisfied, i.e.\ $(3H)^{-1} (dI/dt)_{|{t=t_\star}} = 1$. The region of parameter space in which $2 \leq N(t_\star) < 3$, in this case, is shown in blue in Fig.~\ref{fig:betaNtotsc}. The minimal number of e-folds consistent with two-bubble nucleation is
$N_{\text{tot,min}}^\star = 5.5$.

For this scenario to hold, we assume that the critical temperature $T_c$ lies below the temperature at which vacuum domination sets in, such that the decay cannot begin during radiation domination. Moreover, consistency of the scenario requires that the thermal tunneling rate increases as the temperature drops below $T_c$. For temperatures close to the critical point, $T_2 < T_1 \lesssim T_c$, we need  
\begin{equation}
    \frac{\Gamma_3(T_2)}{\Gamma_3(T_1)}
    \simeq \exp\!\left[(\beta_H-4)\bigl(N_\text{tot}(T_2)-N_\text{tot}(T_1)\bigr)\right] > 1,
\end{equation}
which implies $\beta_H > 4$. Fig.~\ref{fig:betaNtotsc} shows that the parameter space consistent with $2 \leq N(t_\star) < 3$ indeed satisfies this bound, though only marginally. This proximity raises concerns about the reliability of this region.

Using Eq.~\eqref{eq:Rmean} for the mean bubble size, we find that at completion $R_\star(t_\star) H_\star \sim 0.9$ within the green region of Fig.~\ref{fig:betaNtotsc}, making it plausible that bubbles collide when $R_\star H_\star \sim 0.5$. However, in the purple region above the red dashed line, the mean bubble size is  $0.9<R_\star H_\star < 1.8$, and below the dashed line one finds $R_\star H_\star \gg 1$. In this regime a single bubble already spans (or exceeds) the Hubble volume, so the assumption that the transition completes through the collision of two bubbles is no longer self-consistent.

For supercooling to be governed by thermal tunneling, one must verify in any specific model that $\Gamma_3 > \Gamma_4$ up to completion and that the thermal description remains valid, i.e.\ that the thermalization rate satisfies $\Gamma_\text{th} > H^4$.

In conclusion, if quantum tunneling dominates, the transition is expected to end with far more than two bubbles per Hubble volume. When thermal tunneling dominates the supercooled transition, there exists a narrow region of $(\beta_H,\, N_\text{tot}^\star)$ in which the transition completes with an average of two bubbles. However, this region lies very close to the boundary of the consistency conditions, and it is unclear whether realistic models can reliably realize such a scenario.

\end{document}